\documentclass[runningheads]{llncs}
\usepackage{graphicx,amsmath,amssymb}
\begin{document}
\title{cCorrGAN: Conditional Correlation GAN for Learning Empirical Conditional Distributions in the Elliptope}
\titlerunning{Conditional CorrGAN}
%
%
\author{Gautier Marti\inst{1}\orcidID{0000-0001-6497-5702} \and
Victor Goubet\inst{2}\orcidID{0000-0003-4078-4703} \and
Frank Nielsen\inst{3}\orcidID{0000-0001-5728-0726}}
\authorrunning{G. Marti et al.}
%
\institute{Independent researcher\\
\email{} \and
ESILV - \'Ecole Sup\'erieure d'Ing\'enieurs L\'eonard de Vinci, Paris, France
\email{}\\
\url{} \and
Sony Computer Science Laboratories Inc, Tokyo, Japan\\
\email{}}
\maketitle              
\begin{abstract}
We propose a methodology to approximate conditional distributions in the elliptope of correlation matrices based on conditional generative adversarial networks. We illustrate the methodology with an application from quantitative finance: Monte Carlo simulations of correlated returns to compare risk-based portfolio construction methods. 
Finally, we discuss about current limitations and advocate for further exploration of the elliptope geometry to improve results.

\end{abstract}
\section{Introduction}

Since the seminal Generative Adversarial Networks (GANs) paper~\cite{goodfellow2014generative},
 adversarial training has been successful in several areas which are typically explored by machine learning researchers (e.g. image generation in computer vision, voice and music generation in audio signal processing). To a smaller extent, GANs were also applied for generating social and research citations networks.
We are not aware of existing generative adversarial networks for sampling matrices from an empirical distribution defined in the elliptope, the set of correlation matrices, except our previous work \cite{marti2020corrgan}.
In this work, we propose extensions of the original CorrGAN model keeping in mind matrix information geometry~\cite{nielsen2013matrix}.

This body of work can be motivated by applications in quantitative finance, and possibly in other fields relying on correlation and covariance matrices. Having access to a generative model of realistic correlation matrices can enable large scale numerical experiments (e.g. Monte Carlo simulations) when current mathematical guidance falls short.
In the context of quantitative finance, it can improve the falsifiability of statements and more objective comparison of empirical methods. For example, paper A and paper B concurrently claim their novel portfolio allocation methods are the state of the art, out-performing the rest of the literature on well-chosen period and universe of assets.
Can we verify and obtain better insights than these claims?
We will briefly illustrate that the generative models presented in this paper can help perform a more reliable comparison between portfolio allocation methods.
Moreover, by analyzing the simulations, we are able to extract insights on which market conditions are more favorable to such or such portfolio allocation methods.

\section{Related Work}

Research related to our work can be found in three distinct areas: generation of random correlation matrices, generative adversarial networks (in finance), information geometry of the elliptope in machine learning.

\subsection{Sampling of Random Correlation Matrices}

Seminal methods detailed in the literature are able to generate random correlation matrices without any structure. These methods are sampling uniformly (in a precise sense) in the set of correlation matrices, e.g. the onion method \cite{ghosh2003behavior}.
Extended onion methods and vines can be used to generate correlation matrices with large off-diagonal values \cite{lewandowski2009generating}, but they are not providing any other particular structure by design.
Other methods allow to simulate correlation matrices with given eigenvalues \cite{davies2000numerically}, or with the Perron–Frobenius property \cite{huttner2019simulating}.
CorrGAN \cite{marti2020corrgan} is the first and so far only method which attempts to learn a generative model from a given empirical distribution of correlation matrices. Preliminary experiments have shown that the synthetic correlation matrices sampled from the generative model have similar properties than the original ones.


\subsection{GANs in Finance}

Generative Adversarial Networks (GANs) in Finance are just starting to be explored. Most of the research papers (and many are still at the preprint stage) were written after 2019. The main focus and motivation behind the use of GANs in Finance is {\em data anonymization}~\cite{assefa2020generating}, followed by univariate time series (e.g. stock returns) modeling \cite{fu2019time,koshiyama2020generative,takahashi2019modeling,wiese2020quant}.
Also under investigation: modeling and generating tabular datasets which are often the format under which the so-called `alternative data' hedge funds are using is structured; An example of such `alternative data' can be credit card (and other) transactions which can also be modeled by GANs \cite{zheng2018generative}.
To the best of our knowledge, only our previous work \cite{marti2020corrgan} tackles the problem of modeling the joint behaviour of a large number of assets.

\subsection{Geometry of the Elliptope in Machine Learning}

Except a recent doctoral thesis \cite{david2019riemannian}, the core focus of the research community is on the Riemannian geometry of the SPD cone, e.g. \cite{moakher2005differential,barachant2013classification}, SPDNet \cite{huang2017riemannian} and its extension SPDNet with Riemannian batch normalization \cite{NEURIPS2019_6e69ebbf}. Concerning the elliptope, a non-Riemannian geometry (Hilbert's projective geometry) has also been considered for clustering \cite{nielsen2019clustering}.
The Hilbert elliptope distance is an example of non-separable distance which satisfies the information monotonicity property \cite{nielsen2019clustering}.

\section{Our Contributions}

We suggest the use of GANs to learn empirical conditional distributions in the correlation elliptope;
We highlight that the correlation elliptope has received less attention from the information geometry community in comparison to the symmetric positive definite cone of covariance matrices;
We highlight where information geometric contributions could help improve on the proposed models (the Euclidean Deep Convolutional Generative Adversarial Network (DCGAN) is already competitive if not fully satisfying from a geometric point of view);
Finally, we illustrate the use of such generative models with Monte Carlo simulations to understand the empirical properties of portfolio allocation methods in quantitative finance.


\section{The Set of Correlation Matrices}

\subsection{The Elliptope and its Geometrical Properties}

Let $\mathcal{E} = \{C \in \mathbb{R}^{n \times n} ~|~  C = C^\top$, $\forall i \in \{1, \ldots, n\}$, $C_{ii} = 1, \forall x \in \mathbb{R}^{n}$, $x^\top C x \geq 0\}$ be the set of correlation matrices, also known as the correlation elliptope.

\begin{figure}
\centering
\begin{minipage}{.39\linewidth}
    \includegraphics[width=\linewidth]{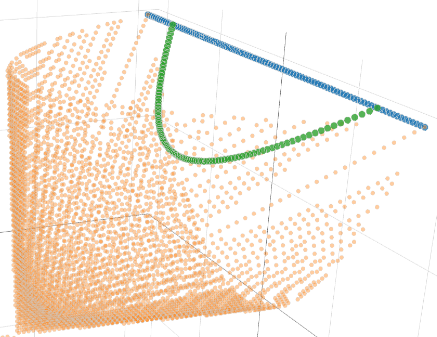}
    \caption{$2 \times 2$ correlation matrices (blue); Fisher-Rao geodesic (green) between two matrices. The geodesic does not stay inside the set of correlation matrices.}
    \label{img1}
\end{minipage}
\hfill
\begin{minipage}{.57\linewidth}
    \includegraphics[width=\linewidth]{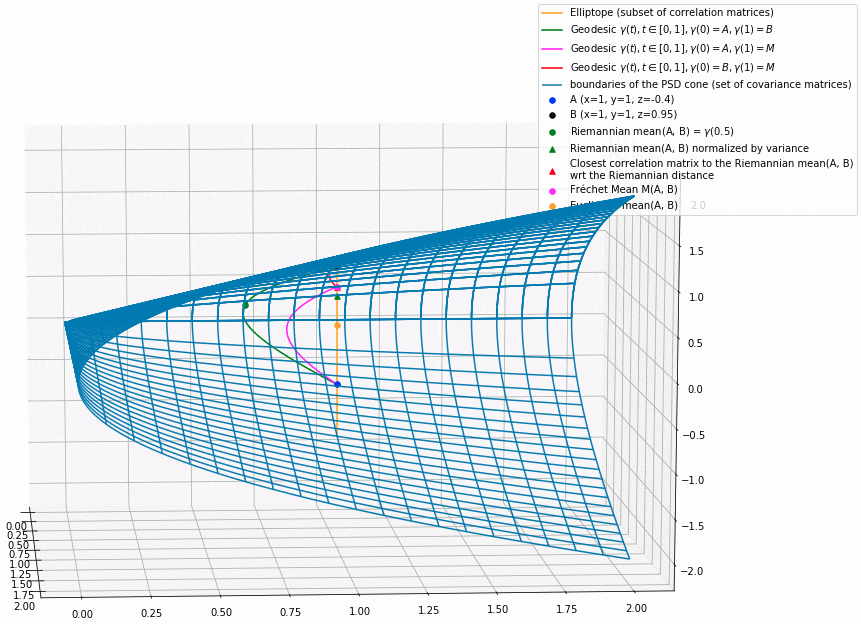}
    \caption{Visualizing various means.}
    \label{img2}
\end{minipage}
\end{figure}

The convex compact set of correlation matrices $\mathcal{E}$ is a strict subspace of the set of covariance matrices whose Riemannian geometry has been well studied in the information geometry literature, e.g. Fisher-Rao distance \cite{wells2020fisher}, Fr\'echet mean \cite{moakher2005differential}, PCA \cite{horev2016geometry}, clustering \cite{shinohara2010covariance,lee2015geodesic}, Riemannian-based kernel in SVM classification \cite{barachant2013classification}.
However, $\mathcal{E}$ endowed with the Riemannian Fisher-Rao metric is not totally geodesic (cf. Figures~\ref{img1},~\ref{img2}), and this has strong implications on the principled use of the tools developed for the covariance matrices.
For example, consider SPDNet \cite{huang2017riemannian} (with an eventual Riemannian batch normalization layer \cite{NEURIPS2019_6e69ebbf}). Despite this neural network is fed with correlation matrices as input, it will generate covariance matrices in subsequent layers. How can we stay in the elliptope in a principled and computationally efficient fashion?

In Figure~\ref{img1}, each $2 \times 2$ covariance matrix is represented by a 3D point $(x, y, z)$.
The blue segment $(x = z = 1)$ is the set of $2 \times 2$ correlation matrices.
In green, the geodesic (using the Riemannian Fisher-Rao metric for covariances) between correlation matrix $\gamma(0) = (1, -0.75, 1)$ and correlation matrix $\gamma(1) = (1, 0.75, 1)$.
The geodesic $t \in [0, 1] \rightarrow \gamma(t)$ (and in particular the Riemannian mean $\gamma(0.5)$) is not included in the blue segment representing the correlation matrix space.

In Figure~\ref{img2} we compare several means: 
\begin{itemize}
    \item (M1) Euclidean mean $\frac{A + B}{2}$,
    \item (M2) Riemannian barycenter (in general, a covariance matrix $\Sigma^\star$),
    \item (M3) $C^\star = \mathrm{diag}(\Sigma^\star)^{-\frac{1}{2}} \Sigma^\star \mathrm{diag}(\Sigma^\star)^{-\frac{1}{2}}$ ($\Sigma^\star$ normalized by variance),
    \item (M4) Fr\'echet mean constrained to the space of correlation matrices,
    \item (M5) projection of $\Sigma^\star$ onto the elliptope using the Riemannian distance.
\end{itemize}
We observe that (M3) is close but not equal to (M4) and (M5), which seems to be equivalent. 
We have not found yet how to compute (M4) or (M5) efficiently. Alternatively, a recent doctoral dissertation \cite{david2019riemannian} on Riemannian quotient structure for correlation matrices might help progress in this direction.



With a better understanding on how to define and efficiently compute a Riemannian mean for correlation matrices, we can think of adapting SPDNet to correlation matrices, and eventually use it as a component of the generative adversarial network architecture.

\subsection{Financial Correlations: Stylized Facts}

Since we aim at learning empirical (conditional) distributions, different domains may have different type of typical correlation matrices whose set describes a subspace of the full correlation elliptope.
In the present work, we will showcase the proposed model on correlation matrices estimated on stocks returns. We briefly describe the known properties, also known as stylized facts, of these correlation matrices. We leave for future work or research collaborations to explore the empirical space of correlation matrices in other domains.

The properties of financial correlation matrices are well-known and documented in the literature.
However, until CorrGAN \cite{marti2020corrgan}, no single model was able to generate correlation matrices verifying a handful of them at the same time: (SF 1) Distribution of pairwise correlations is significantly shifted to the positive; (SF 2) Eigenvalues follow the Marchenko–Pastur distribution, but for a very large first eigenvalue; (SF 3) Eigenvalues follow the Marchenko–Pastur distribution, but for $\approx 5\%$ of other large eigenvalues; (SF 4) Perron-Frobenius property (first eigenvector has positive entries); (SF 5) Hierarchical structure of correlation clusters; (SF 6) Scale-free property of the corresponding Minimum Spanning Tree.



When markets are under stress, it is widely known among practitioners that the average correlation is high.
More precisely, the whole correlation structure changes and becomes different than the ones typical of rallying or steady markets.
These characteristics tied to the market regime would be even harder to capture with a fully specified mathematical model.


\section{Learning Empirical Distributions in the Elliptope using Generative Adversarial Networks}

We suggest that conditional generative adversarial networks are promising models to generate random correlation matrices verifying all these not-so-well-specified properties.

\subsection{CorrGAN: Sampling Realistic Financial Correlation Matrices}

The original CorrGAN \cite{marti2020corrgan} model is essentially a DCGAN (architecture based on CNNs).
CNNs have 3 relevant properties which align well with stylized facts of financial correlation matrices: shift-invariance, locality, and compositionality (aka hierarchy). Based on our experiments, this architecture yields the best off-the-shelf results most consistently.
However, outputs (synthetic matrices) need to be post-processed by a projection algorithm (e.g. \cite{higham2002computing}) as they are not strictly valid (but close to) correlation matrices. This goes against the end-to-end optimization of deep learning systems.
Despite CorrGAN yields correlation matrices which verify the stylized facts, we can observe with PCA (or MDS, t-SNE) that the projected empirical and synthetic distributions do not match perfectly. We also noticed some instability from one trained model to another (well-known GAN defect). For example, in our experiments, the average Wasserstein distance (using the POT package \cite{flamary2017pot}) between any two (PCA-projected) training sets is $\mu_E := 6.7 \pm \sigma_E := 6.8$ (max distance := 15) whereas the average distance between a training set and a generated synthetic set is $\mu_G := 18.8 \pm \sigma_G := 8$ (min distance := 8). Ideally, we would obtain $\mu_E \approx \mu_G$ instead of $\mu_E \ll \mu_G$. We observed that CorrGAN was not good at capturing the different modes of the distribution. This motivated the design of a conditional extension of the model.




\subsection{Conditional CorrGAN: Learning Empirical Conditional Distributions in the Elliptope}

\begin{figure}
\centering
\begin{minipage}{.52\linewidth}
    \includegraphics[width=\linewidth]{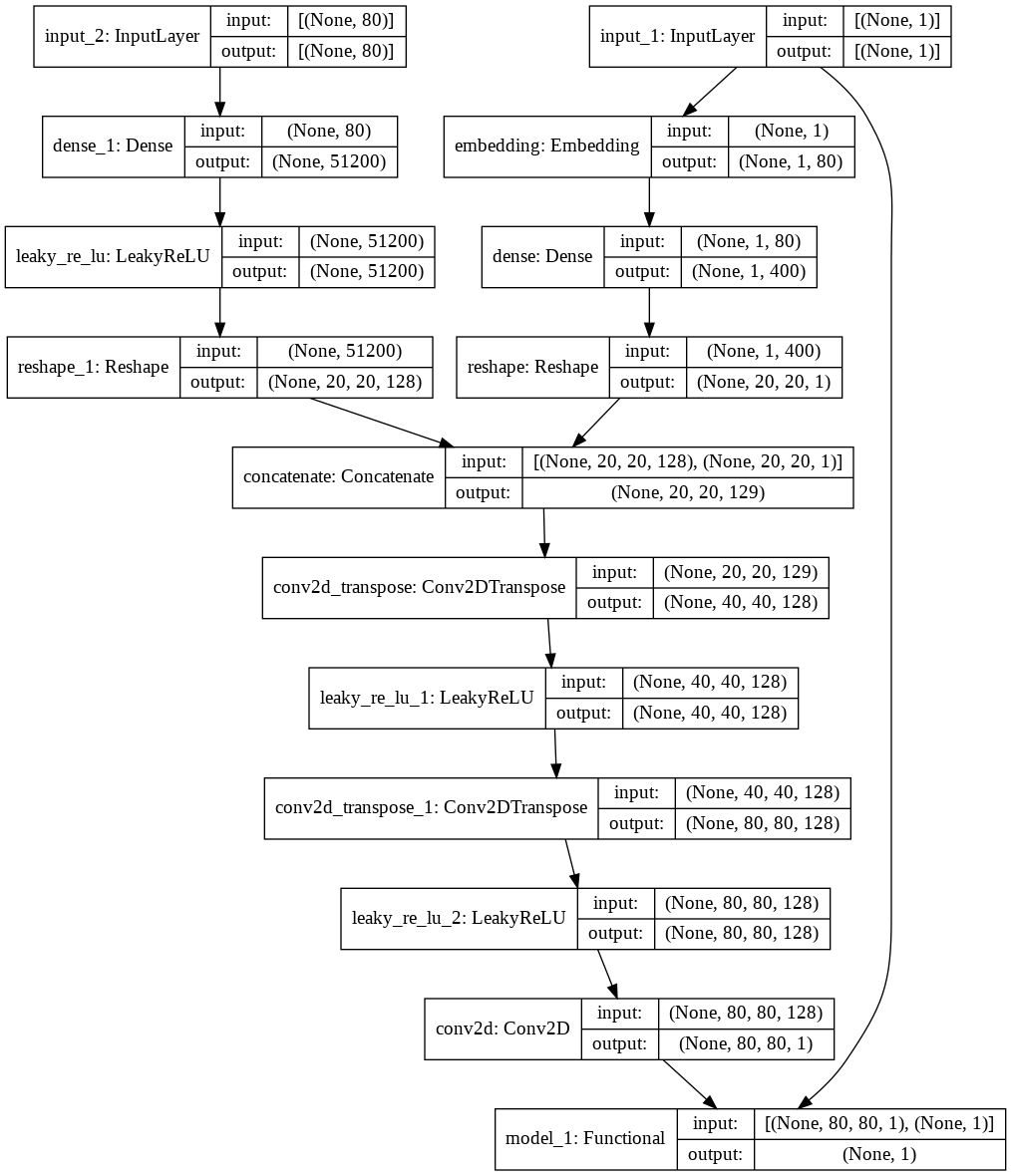}
    \caption{Architecture of the conditional GAN (cCorrGAN) which generates $80 \times 80$ correlation matrices corresponding to a given market regime $\{$\texttt{stressed}, \texttt{normal}, \texttt{rally}$\}$}
    \label{img3}
\end{minipage}
\hfill
\begin{minipage}{.44\linewidth}
    \includegraphics[width=\linewidth]{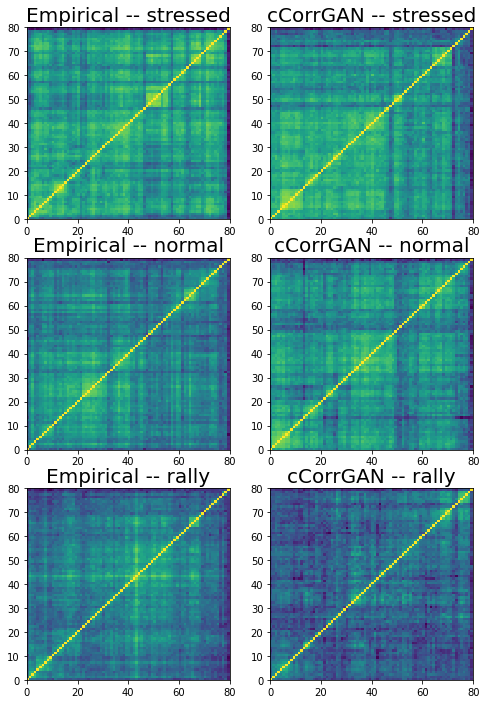}
    \caption{Empirical (left) and cCorrGAN generated (right) correlation matrices for three market regimes}
    \label{img4}
\end{minipage}
\end{figure}

We describe in Figure~\ref{img3} the architecture of cCorrGAN, the conditional extension of the CorrGAN model \cite{marti2020corrgan}.
To train this model, we used the following experimental protocol: 1. We built a training set of correlation matrices labeled with one of the three classes $\{$\texttt{stressed}, \texttt{normal}, \texttt{rally}$\}$ depending on the performance of an equally weighted portfolio over the period of the correlation estimation; 2. We trained several cCorrGAN models; 3. We evaluated them qualitatively (PCA projections, reproductions of stylized facts), and quantitatively (confusion matrix of a SPDNet classifier \cite{NEURIPS2019_6e69ebbf}, Wasserstein distance between projected empirical and synthetic distributions) to select the best model. Samples obtained from the best model are shown in Figure~\ref{img4}.


\section{Application: Monte Carlo simulations of risk-based portfolio allocation methods}

\begin{figure}
\includegraphics[width=\textwidth]{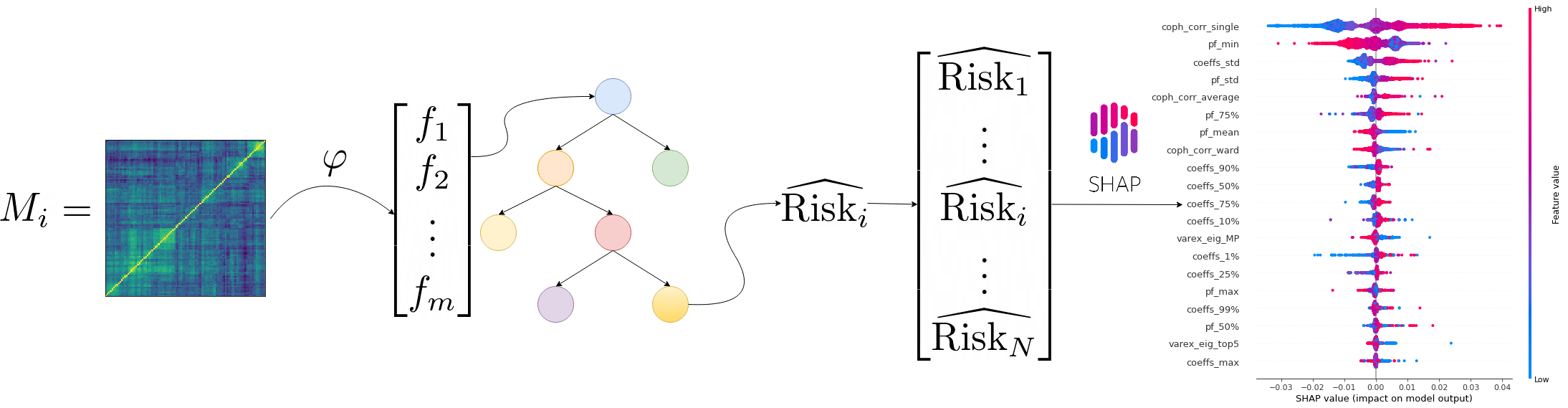}
\caption{Methodology for analysing the results of the Monte Carlo simulations} \label{fig3}
\end{figure}

Monte Carlo simulations to test the robustness of investment strategies and portfolio construction methods were the original motivation to develop CorrGAN, and its extension conditional CorrGAN.
The methodology is summarized in Figure~\ref{fig3}: 1. Generate a random correlation matrix from a given market regime using cCorrGAN, 2. extract features from the correlation matrix, 3. estimate the in-sample and out-of-sample risk on synthetic returns which verify the generated correlation structure, 4. predict and explain (with Shapley values) a) performance decay from in-sample to out-of-sample, b) out-performance of a portfolio allocation method over another.
The idea of using Shapley values to explain portfolio construction methods can be found in \cite{papenbrock2020matrix}.



Using such methodology, enabled by the sampling of realistic random correlation matrices conditioned on a market regime, we can find, for example, that the Hierarchicaly Risk Parity (HRP) \cite{de2016building} outperforms the naive risk parity when there is a strong hierarchy in the correlation structure, well-separated clusters, and a high dispersion in the first eigenvector entries, typical of a normal or rallying market regime. In the case of a stressed market where stocks plunge altogether, HRP performs on-par (or slightly worse because of the risk of identifying a spurious hierarchy) than the simpler naive risk parity.








\vskip 0.3cm
\noindent {\bf Acknowledgments:}
{\small
We thank ESILV students for helping us running experiments for hundreds of GPU-days: Chlo\'e Daems, Thomas Graff, Quentin Bourgue, Davide Coldebella, Riccardo Rossetto, Mahdieh Aminian Shahrokhabadi, Marco Menotti, Johan Bonifaci, Cl\'ement Roux, Sarah Tailee, Ekoue Edem Ekoue-Kouvahey, Max Shamiri.}

\bibliographystyle{splncs04}
\bibliography{biblio}

\end{document}